\documentclass[a4paper]{article}

\usepackage{INTERSPEECH2020}

\usepackage[table,xcdraw]{xcolor}
\usepackage[font=small,skip=5pt]{caption}
\usepackage{multicol}
\usepackage{multirow}
\usepackage{tabularx, makecell}
\usepackage{graphicx}
\usepackage{subcaption}
\title{Correlation based Multi-phasal models for improved imagined speech EEG recognition}

\name{Rini A Sharon$^1$, Hema A Murthy$^1$}
\address{
  $^1$Indian Institute of Technology, Madras} 
\email{ee15d210@smail.iitm.ac.in, hema@cse.iitm.ac.in}

\begin{document}

\maketitle
\begin{abstract}
Translation of imagined speech electroencephalogram(EEG) into human understandable commands greatly facilitates the design of naturalistic brain computer interfaces. To achieve improved imagined speech unit classification, this work aims to profit from the parallel information contained in multi-phasal EEG data recorded while speaking, imagining and performing articulatory movements corresponding to specific speech units. A bi-phase common representation learning module using neural networks is designed to model the correlation and reproducibility between an analysis phase and a support phase. The trained Correlation Network is then employed to extract discriminative features of the analysis phase. These features are further classified into five binary phonological categories using machine learning models such as Gaussian mixture based hidden Markov model and deep neural networks. The proposed approach further handles the non-availability of multi-phasal data during decoding. Topographic visualizations along with result-based inferences suggest that the multi-phasal correlation modelling approach proposed in the paper enhances imagined-speech EEG recognition performance. 


\end{abstract}
\noindent\textbf{Index Terms}: Speech-EEG recognition, Brain-computer interface, Correlation Networks

\section{Introduction}

Brain Computer Interfaces(BCI) are devices that collect brain signals and translate them into human-understandable commands\cite{schalk2004bci2000}. Such interfaces facilitate communication impaired individuals to interact with the external world thereby improving their standard of living\cite{lazarou2018eeg, soman2015using}. The brain signals used to control BCIs can be collected using invasive, semi-invasive or non-invasive methods. Invasive and semi-invasive techniques like deep-brain recording and electrocorticography  require electrodes to be inserted directly into the human brain or skull by surgical interventions\cite{hill2012recording, nardone2014invasive}. Conversely non-invasive methods like Electroencephalography(EEG) capture the electrical activity occurring in the brain by employing electrodes that measure scalp potentials.  Although EEG signals suffer from low signal-to-noise ratio, poor spatial resolution and artifact dominance, their high temporal resolution, ease-of-use and cost-effective nature makes them a popular recording alternative\cite{zion7eeg, beres2017time}.

Despite the numerous applications of decoding brain signals to establish communication,  a majority of existing BCI modules are based on non-speech inputs like mental counting\cite{wang2013study,altiindaug2017discover}, imagined motor movements\cite{alimardani2018brain, gonzalez2011motor} and visual/auditory evoked potentials\cite{wang2008brain, kidmose2013study}. A more naturalistic design would involve the subject using speech based inputs as controls to the BCI devices\cite{riniembc,rinincc}. There exist different speech-associated cognitive phases that can be chosen as controls. These include attention-based audition, speech production, verbal miming, articulatory movements, speech imagination and so on performed by the subjects\cite{anumanchipalli2019speech,sharon2019empirical,riniaccess}.

Among these phases, speech imagination finds potential application as a BCI control\cite{mohanchandra2015eeg, d2009toward}. Imagined speech BCI is highly advantageous for individuals who are unable to move their articulators because of physical disabilities such as locked-in syndrome,
or advanced amyotrophic lateral sclerosis\cite{brigham2010imagined}. Likewise, in military applications and gaming applications where visual or audible communication is undesirable, silent speech is an effective and safer alternative. The convenience of usage and practical applicability of EEG-based imagined speech BCIs have made them popular\cite{neubig2019recognition, hashim2017word, brigham2010imagined}. However the accuracies reported in the field suggest significant scope for improvement with regard to the design of signal processing and decoding modules. 

Our work is a novel attempt to improve imagined speech EEG recognition by constructing a feature extraction module that utilizes inter-phasal information during the training phase to strengthen single phase unit discrimination. Section \ref{sec:motiv} briefly discusses the motivation behind this work and how it compares with other similar domain works. The proposed framework is detailed in Section \ref{sec:prof} following which the experimental details are outlined in Section \ref{sec:exp}. Section \ref{sec:result} discusses the results and inferences of this work and Section \ref{sec:conc} summarizes the paper.


\section{Motivation and Related Work}\label{sec:motiv}

Multiple works on silent speech recognition deal with uni-modal single-phase imagined speech EEG(is-EEG) of vowels, syllables or words\cite{brigham2010imagined, manca2016eeg, imani2017ica}. Recently multi-modal and multi-phasal data inclusion has proved to be beneficial in boosting is-EEG recognition rates\cite{zhao2015classifying, sun2016neural}. Multi-modal information includes recording facial video and speech audio alongside EEG and multi-phasal refers to the EEG data recorded during different phases of subject activities like speech production, imagination or articulation. In addition to speech audio, \cite{sun2016neural} also uses bi-phasal data corresponding to spoken-EEG and is-EEG simultaneously to decode is-EEG and provides proof of correlation between these phases by visualizing their spectrum based scalp distributions. However, the equipment requirements for multi-modal data recordings, and their syncing lead to inconvenience of deployment in real-time BCIs. Furthermore, recognizing is-EEG during test time would require the multi-modal or multi-phasal information to be present, further delaying the decode duration. Considering these drawbacks and the evidence of inter-phasal correlation\cite{sun2016neural}, a realistic alternative would be to utilize the multi-phasal information during training and test only on imagined speech EEG. 

Common representation learning(CRL) aims to integrate information from multiple views/modes of the data and project them to a common space\cite{huang2017cross}. Improving single mode testing performance by using data from other modes during training is one of the applications that motivate the importance of CRL\cite{wang2015deep}. Correlation Network(CorrNet) based CRL approaches achieve the same\cite{rinicorrelation, chandar2015correlational}. CorrNets are trained to implicitly maximize the correlation between the views in a common projected space while maintaining balanced self and cross reconstruction accuracies within the views. 


In this paper, we intend to employ CorrNets as feature extraction modules to determine if different phases of EEG data contribute informatively towards better imagine-phase EEG recognition. For this purpose, we analyze an EEG dataset that deals with three speech-based cognitive phases, namely, articulation, speech production and speech imagination. The rest of the paper assumes these phases to be the ``views" for CRL. 

\section{Proposed Framework}\label{sec:prof}

The proposed technique involves two levels of feature extractions, namely, a preliminary handcrafted feature extraction(primary features) followed by a CRL based CorrNet feature extraction(secondary features). CorrNets as secondary feature extractors stand to benefit from the multi-phasal EEG data recorded. This feature extractor is then coupled with machine learning classification modules to perform is-EEG recognition. Since our focus is in decoding imagined speech EEG, this will be referred to as the analysis phase EEG($E_A$). The other phases are considered as the support phases($E_S$), as these features support the analysis phase by improving unit-level discrimination. 

\subsection{Model Architecture and Details}

\begin{figure}[t]
  \centering
  \includegraphics[width=\linewidth,trim={0cm 0cm 0cm 0cm},clip]{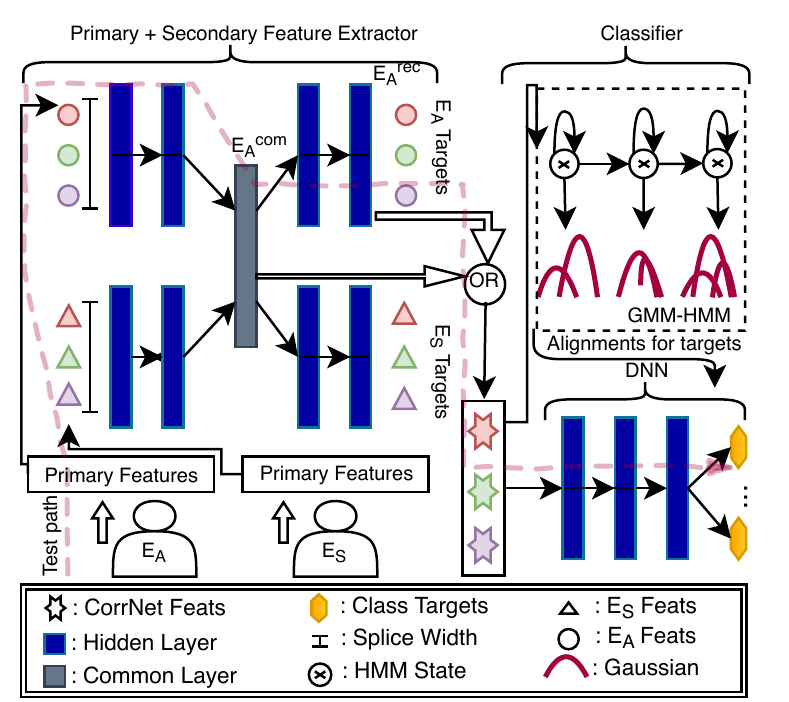}
  \caption{Proposed Architecture; Testing path(dashed line)}
  \label{fig:cor_arc}
  \vspace{-0.6cm}
\end{figure}

After pre-processing and preliminary feature extraction, the CorrNet feature extractor is provided with the EEG features from two input phases, namely $E_A$ and $E_S$ as outlined in Figure \ref{fig:cor_arc}. CorrNets are designed to achieve the combined objective of two popular techniques, canonical correlation analysis(CCA) and multi-modal auto encoders(MAE). CCA aims to learn highly correlated common representations of the different phases of the data\cite{thompson2005canonical}. MAE on the other hand intends to perform self reconstruction and cross reconstruction of the phases of the data\cite{jaques2017multimodal}. Bringing together these objectives, CorrNet is trained to maximize the correlation between the phases in a common projected space as well as achieve good cross and self-reconstruction.

The CorrNet architecture thus comprises of a bottleneck common shared layer sandwiched between phase-specific input and output layers. The input layer and output layers have the same number of units and differ from the number of shared hidden layer units. The input layer projects the two phases in a common space in such a way that the correlation between their projections is maximized. The hidden layer computes an encoded representation of the input and the output layer tries to reconstruct the input from this hidden representation. Hence the overall training optimization loss function for two input phases $E_A$ and $E_S$ includes the following terms:

\begin{enumerate}
    \item $\mathbb{L}_a=\text{Mean Squared Error}([E_A, \_], E_A^{rec})$
    \item $\mathbb{L}_b=\text{Mean Squared Error}([\_, E_S], E_A^{rec})$
    \item $\mathbb{L}_c=\text{Mean Squared Error}([E_A, E_S], E_A^{rec})$
    \item $\mathbb{L}_d=\text{Correlation Loss}(\text{Projected}(E_A), \text{Projected}(E_S)])$
\end{enumerate}
\begin{equation}
   \textrm{Corr-Loss(A,B)}  = \frac{\sum_{i=1}^{N}(A_{i}-\overline{A})(B_{i}-\overline{B})}{\sqrt{\sum_{i=1}^{N}(A_{i}-\overline{A})^2\sum_{i=1}^{N}(B_{i}-\overline{B})^2}} 
  \label{eq1}
  \vspace{-0.5cm}
\end{equation}

\begin{equation}
\textrm{Overall Loss} = \alpha \mathbb{L}_a  + \beta \mathbb{L}_b + \gamma \mathbb{L}_c - {} \lambda \times \mathbb{L}_d 
\label{eq2}
\end{equation}
where Mean Squared Error($[in1, in2], rec$) is said to denote the loss when the CorrNet is provided with inputs ``in1" as view-1 and ``in2" as view-2. This loss is calculated between the original $E_A$ features and their reconstructed versions denoted by the superscript ``rec". The correlation loss(corr-loss:$\mathbb{L}_d$) is calculated between the projections of the two phases in the shared layer for $N$ input training samples as given in Equation \ref{eq1}. In Equation \ref{eq2}, $\alpha$, $\beta$ and $\gamma$ are the weights given to the losses while training to emphasise/de-emphasise the losses and analyse their effects on the overall feature classification. The scaling factor $\lambda$ is used to adjust the range of the corr-loss to match that of the reconstruction losses. $\lambda$ is preceded with a negative sign because while we want to minimize the reconstruction losses, we want to maximise correlation loss. In other words, we state that the CorrNet is trained and optimized to perform the following:
\begin{itemize}
    \item Reconstruct $E_A$ from itself (Self-reconstruction)
    \vspace{-0.1cm}
    \item Reconstruct $E_A$ from $E_S$ (Cross-reconstruction)
    \vspace{-0.1cm}
    \item Reconstruct $E_A$ when both $E_A$ and $E_S$ are provided as inputs (Joint-reconstruction)
    \vspace{-0.1cm}
    \item Maximize the correlation between $E_A$ and $E_S$ in the common projected space
\end{itemize}
   
\vspace{-0.3cm}

\subsection{Training and Testing methodology}

The experimental structure is divided into two stages, namely, the training stage which contains the multi-phasal data and the testing stage which provides the analysis phase data alone. Since a bi-phasal framework is considered for modelling the CorrNet, $E_A$ and $E_S$ will be fed as inputs as well as targets to the CorrNet during training. On completion of training, the input instances of $E_A$ are projected through the CorrNet. Both the common layer projections($E_A^{com}$) and the reconstructed analysis phase projections($E_A^{rec}$) are considered as outputs of the feature extraction module as marked in Figure \ref{fig:cor_arc}. A performance-based choice is made between these projected features and the same is implemented to train the final classification module.  

In the course of testing, $E_A$ trials are passed through the CorrNet feature extractor and the pre-chosen projection path features are computed. These are then passed to the trained classifier to obtain is-EEG predictions.

\begin{figure}[t]
  \centering
  \includegraphics[width=\linewidth,trim={0cm 6cm 7cm 0cm},clip]{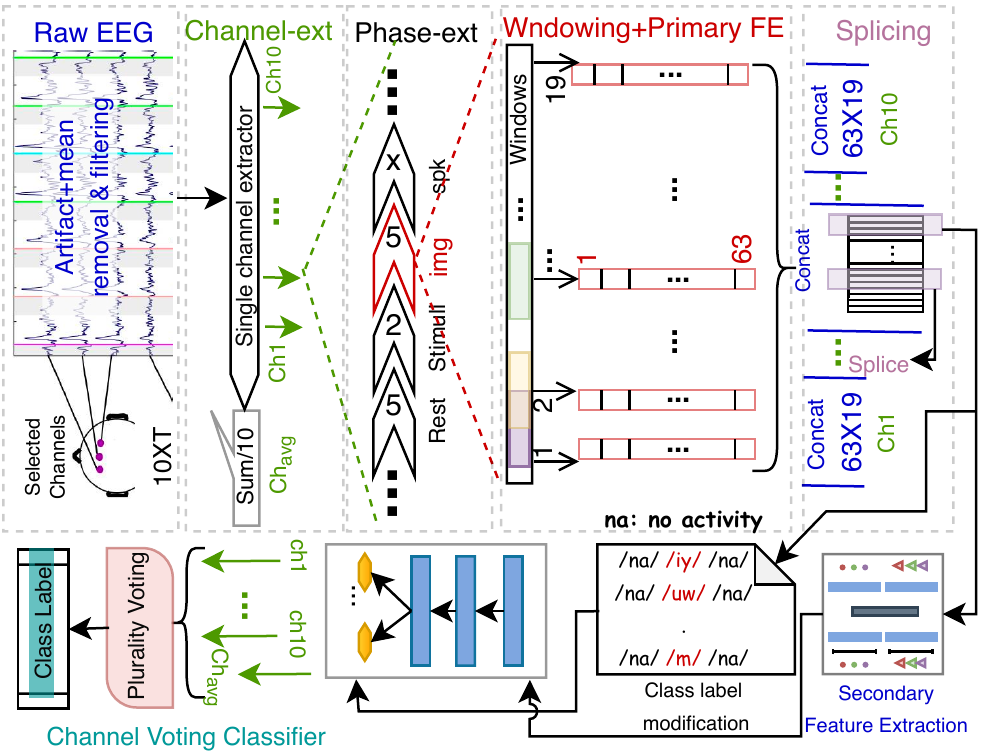}
  \caption{Flow diagram describing how Raw EEG is processed, primary and secondary features are extracted and classified}
  \vspace{-0.1cm}
  \label{fig:flow}
  \vspace{-0.6cm}
\end{figure}

\section{Experiments}\label{sec:exp}
All the experiments in this work consider a leave-one-subject out testing scheme and report average classification accuracies over the 14 subjects in the Database. A grid search based tuning of architectural parameters and hyper-parameters was performed to obtain an optimal set of design parameters. The experimental flow is pictorially depicted in Figure \ref{fig:flow}.

\subsection{Dataset}
The publicly available KARA ONE database which consists of multi-modal and multi-phasal EEG recordings has been considered to test our proposed models(details available in \cite{zhao2015classifying}).
A 64-channel Neuroscan cap was used to record the EEG data at a sampling rate of 1kHz using an electrode placement that follows the 10-20 system. The subjects were instructed to minimize movements and the cues were presented in a computer screen. 4 words(pat, pot, knew, and gnaw) and 7 syllabic(iy, uw, piy, tiy, diy, m, n) prompts were presented to 14 subject, 12 times each while their EEG data is recorded. The EEG data collection protocol involved a rest state, a stimulus state involving articulatory movements required to start pronouncing the prompt, an imagined speech state and a spoken speech state per trial. Aligning with the objective of this paper, we discard the multi-modal facial and audio related information and focus on the multi-phasal spoken, imagined and articulated EEG signals.

\subsection{Pre-processing}
Raw EEG signal pre-processing steps similar to those followed in \cite{zhao2015classifying} were adopted for this work. Ocular artifact removal using blind source separation, 1-50Hz bandpass filtering and mean removal was performed.  Trials and states of interest are extracted from the EEG data. From previous works dealing with the same dataset, 10 channels having higher correlations with speech activities were extracted and used for analysis - FC6, FT8, C5, CP3, P3, T7, CP5, C3, CP1, and C4\cite{zhao2015classifying, sun2016neural}.

\subsection{Primary Feature Extraction}
The phase level segmented EEG data is of the dimension 62$\times \mathbb{T}$ where $\mathbb{T}$ is the duration of the segment and 62 channels excluding M1 and M2 are considered. Data corresponding to the 10 analysis channels are extracted to form a 10$\times \mathbb{T}$ data matrix which forms the input of the pre-processing module outlined in Figure \ref{fig:flow}. For every phase in each of these 10 channels, MATLAB is used to extract windowed handcrafted features identical to \cite{zhao2015classifying} but with horizontal concatenation of the windowed features to preserve the temporal structure. Window length is taken to be 0.1 times the length of the segment and a 50\% overlap is considered for successive windows. Since the recording duration of is-EEG is fixed to 5 seconds, windowing gives rise to 19 temporal segments. 21 statistical analysis measures as listed in \cite{zhao2015classifying} were extracted along with their first and second derivatives giving a 63 dimensional feature vector per windowed segment. These are temporally concatenated to form primary features of dimension 10$\times$63$\times$19. For speech EEG and articulatory EEG, the primary feature dimension is 10$\times$63$\times \mathbb{D}$, where $\mathbb{D}$ is the windowed duration of their corresponding segments. 

\subsection{CorrNet Design}
The CorrNet architecture was implemented using the Tensorflow toolkit\cite{abadi2016tensorflow}. A 3-frame temporal context is given as the input to the CorrNet by splicing the input features. A disjoint split of 10\% of the train instances is used as a validation set to tune the network parameters(parameters that provided best discrimination are reported below). Mean and variance normalization is applied to the input features. Two phase specific input and output layers of 50 units each were used in the network and the common layer contained 25 neurons. The loss scaling factor, $\lambda$, was tuned and set to 2.5, and the loss weights were set at $\alpha:1.5$, $\beta:0.5$ and $\gamma:1$. Higher $\alpha$ value places more significance on the test case where only analysis phase is available. ReLU activation and Adam optimizer is used with a batch size of 256 to train the CorrNet till convergence. The learning rate was set to 0.005 and was reduced by a factor of 2 when the performance saturates.  Either the reconstructed analysis phase or the hidden layer projections were considered as feature outputs.

\subsection{Classification Module}
Machine Learning motivated Gaussian Mixture Model(GMM) based Hidden Markov Modelling(HMM) and Deep Neural Networks(DNN) were implemented in Kaldi Toolkit as  classification modules\cite{kipyatkova2016dnn}. GMMs represent the means and variances of the data points and HMMs use states with transition probabilities to model the temporal structure of the data. 5 gaussian mixtures per state is considered in a tri-state HMM. The alignments generated using the GMM-HMM models are used as target alignments for the DNN model. 

A key difference in implementation between the previous works in KARA ONE and our work is the inclusion of non-activity state. Since a fixed duration of 5 seconds is assumed for the is-EEG, temporal modelling using just the prompt label disregards the temporal location of the imagining activity. Hence we design a class label modification framework where the activity region is preceded and succeeded by a non-activity state label(comparable to silence in speech) as shown in Figure \ref{fig:flow}. Once the target labels are modified, the network is trained to perform phonological classification and the accuracies are compared with the baseline and the previous works. 


Primary features corresponding to every channel are passed as separate training instances to the CorrNet feature extractor and the classifier. The average across all channels is also computed and passed as a training instance. The classification module then assigns a class label to each channel's segment. Plurality voting is then performed on the channel predictions across the 11 instances(10 channels + channel-average), to determine the final class label for the segment. 

\begin{table}[b] 
\vspace{-0.1cm}
\caption{Performance comparison with existing work on KARA ONE database; nS: number of subjects considered for analysis}
\label{tab:tab1}
\begin{tabular}{ccccccc}
\rowcolor[HTML]{C0C0C0} 
\textbf{Method}                  & \textbf{nS} & \textbf{Bilab} & \textbf{Nasal} & \textbf{C/V} & \textbf{/uw/} & \textbf{/iy/} \\
\cellcolor[HTML]{EFEFEF} \cite{zhao2015classifying}     & 8                 & 56.64             & 63.5           & 18.08        & 79.16         & 59.6          \\
\cellcolor[HTML]{EFEFEF} \cite{sun2016neural}     & 14                & 53                & 47             & 25           & 74            & 53            \\
\cellcolor[HTML]{EFEFEF} \cite{saha2019speak}     & 14                & 75.55             & 73.45          & 85.23        & 81.99         & 73.30         \\
\cellcolor[HTML]{EFEFEF} \cite{bakhshali2020eeg}     & 8                 & 69.08             & 72.10          & 86.52        & \textbf{83.98}         & 75.27         \\
\rowcolor[HTML]{FFCCC9} 
BL($E_A$)     & 14                 & 71.21             & 71.12          & 76.52        & 79.55         & 73.48        \\
\rowcolor[HTML]{FFCCC9} 
BP($E_A$) & 14                & \textbf{75.64}                & \textbf{76.5}             & \textbf{89.39}           & 82.58            & \textbf{80.30} \\
\rowcolor[HTML]{DAE8FC} 
$E_A$+spk & 14                & 75.77                & 77.82             & 91.5           & 83.8            & 81.6 \\
\rowcolor[HTML]{DAE8FC} 
$E_A$+art & 14                & 75.72                & 76.93             & 90.91           & 83.85            & 81.42
\end{tabular}
\vspace{-0.1cm}
\end{table}

\begin{figure}[t]
  \centering
  \includegraphics[width=\linewidth,trim={0.7cm 0.7cm 0.5cm 2cm},clip]{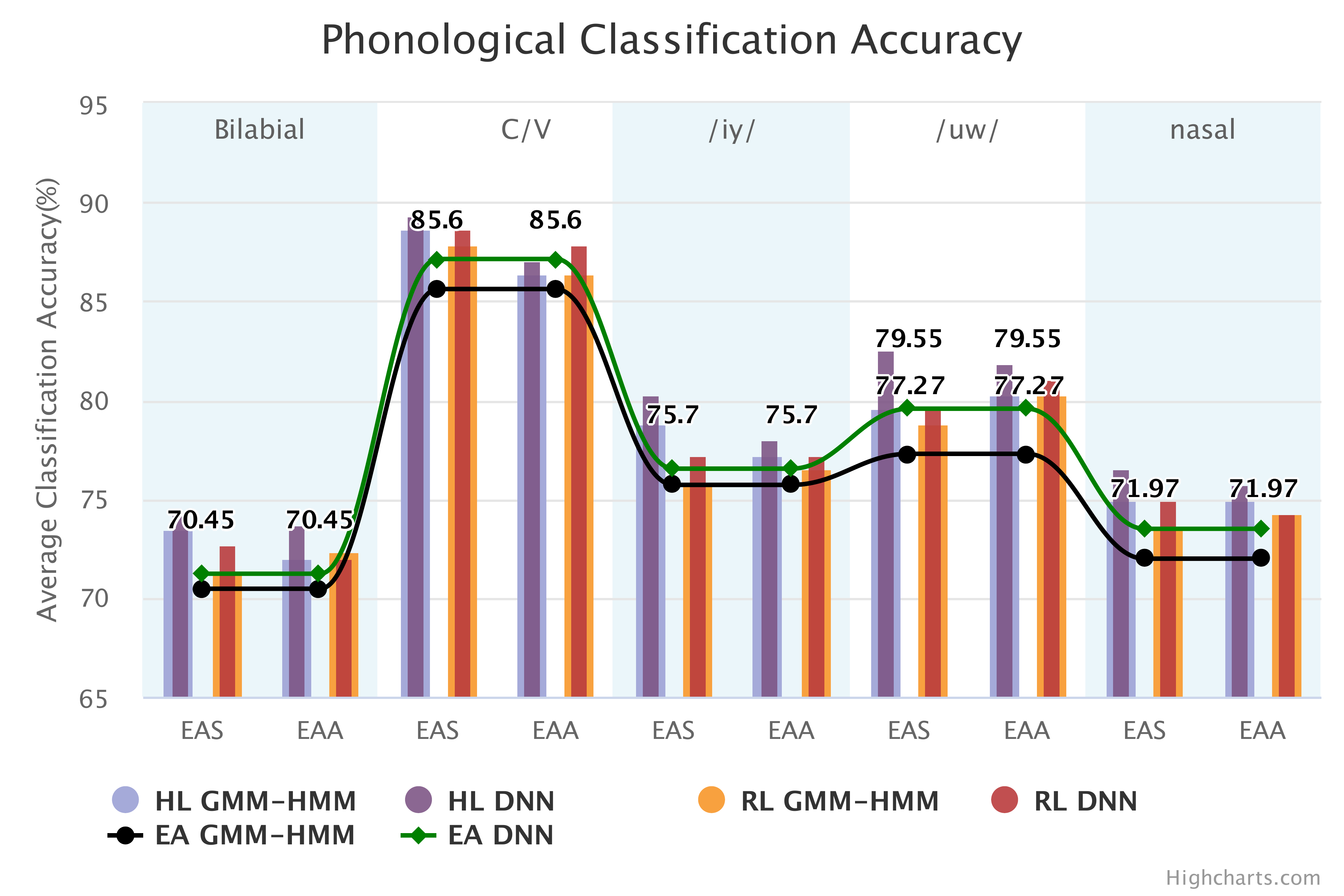}
  \caption{Model Baseline Vs Proposed method performances}
  \label{fig:res}
  \vspace{-0.4cm}
\end{figure}
\section{Results and Discussion}\label{sec:result}
The aim of classifying phonological categories, as in previous works using KARA ONE database, is adopted to evaluate our model. Five categories, namely, presence or absence of bilabial, nasal, $/uw/$, $/iy/$ and consonant+vowel/vowel-only(C/V) are considered in the binary classification framework. 

\subsection{Baseline Vs Proposed Method}
The proposed approach contains multiple modules involving methodological modifications. To examine the effect of every module on the overall classification performance and for the purpose of contrast, we define a baseline(BL) system as a model performing $E_A$(analysis-phase) recognition using primary features as inputs to GMM-HMM/DNN classifier. Since the secondary feature extraction stage has access to two support phases, speaking-EEG and articulatory-EEG, we symbolically represent these using $E_{AS}$ and $E_{AA}$ respectively. The CorrNet output feature projections are taken from either the hidden layer(HL) or the reconstruction layer(RL).  

The classification accuracies of $E_{AS}$ and $E_{AA}$ are compared with our model baseline $E_A$ and reported in Figure \ref{fig:res}. Since the CorrNet output feature projections are taken from either the HL or the RL corresponding to the analysis phase, performances of both these configurations are plotted. We observe that inclusion of CorrNet based secondary feature extractor improves the classification accuracy consistently for all cases across all subjects(absolute average increase of $\approx$3\%). The HL projections capture better discrimination as compared to RL projections. This could result from the increased influence of the support phase correlation in the HL as opposed to the effect of weight-matrix multiplication for reconstruction in RL. Among all the models, HL-projected CorrNet features of the analysis+spoken phase EEG, coupled with a GMM-HMM aligned DNN classifier performs best. A 11-unit is-EEG set-up, trained and tested using this best performing model to classify the 11 prompts in the database yielded a leave-one-subject out average accuracy of 35.82\%. 

\subsection{Comparative Analysis}
KARA ONE is a popularly studied database primarily due to its multi-model/phasal structure. In Table \ref{tab:tab1} we compare our BL and the best performing(BP) model accuracies with those reported earlier(although number of subjects considered differs). With an exception of the $/uw/$ classification case, our model reports the highest classification accuracies. Apart from performance evaluation, another advantage offered by the proposed method as opposed to previous multi-modal contributions\cite{zhao2015classifying,sun2016neural} is the ease and speed of decoding by disregarding the multi-modal/phasal requirement during testing. Assuming we have access to multi-phase information during decode, the model performances are reported in the blue rows of Table \ref{tab:tab1}. Although these provide better classification accuracies, our interest lies in decoding using only the analysis phase information. 

\subsection{Topographic maps}
Inter-phasal correlations were previously visualized topographically for imagined and spoken EEG\cite{sun2016neural}. Here we visualize the trial-averaged topographic plots for the prompt $/n/$ of all three raw EEG phases under consideration without any filtering. The contour patterns of the imagined and speaking phase closely resemble, thus justifying the performance gain obtained while using spoken EEG as a support phase. Although is-EEG is distinct from articulatory EEG in terms of scalp distribution potentials, result-based inferences suggest that articulatory-EEG enhances model discrimination ability. Articulatory-EEG topomap is however similar to spoken-EEG topomap as in all prompts(expect word prompts) absence of voicing is the only major factor distinguishing the two phases.    

\begin{figure}[t]
\begin{subfigure}{0.15\textwidth}
\centering
\includegraphics[width=\linewidth, trim={2cm 1.5cm 2cm 1cm},clip]{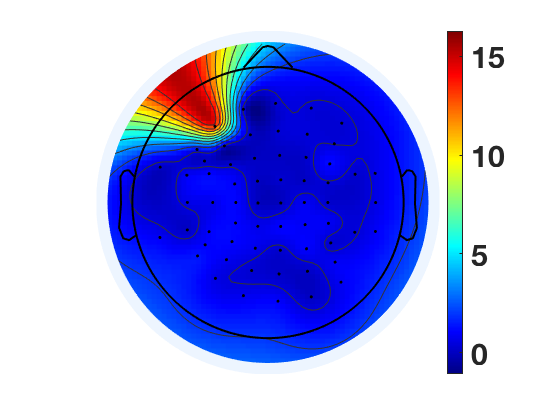}\quad
\caption{Imagination}
  \label{fig:11}
\end{subfigure}
\begin{subfigure}{0.15\textwidth}
\centering
\includegraphics[width=\linewidth, trim={2cm 1.5cm 1.7cm 1cm},clip]{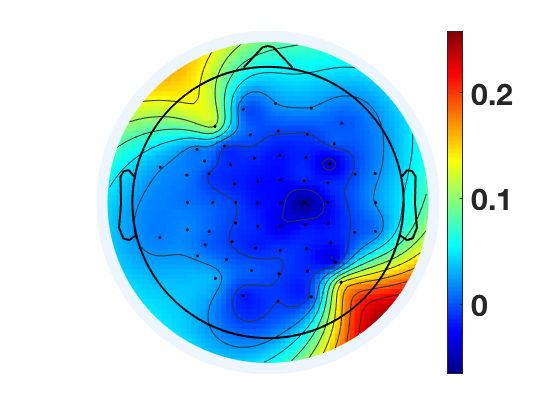}
\caption{Production}
  \label{fig:11}
\end{subfigure}
\begin{subfigure}{0.15\textwidth}
\centering
\includegraphics[width=\linewidth, trim={2cm 1.5cm 1.6cm 1cm},clip]{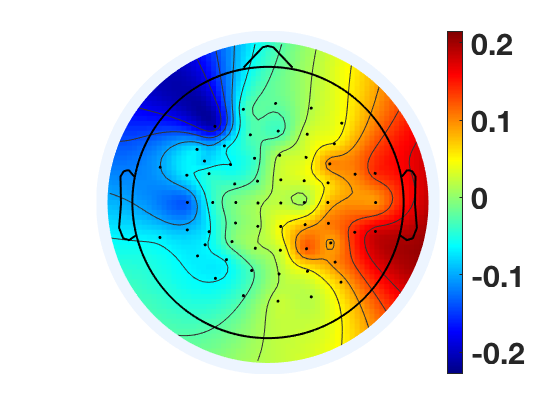}
\caption{Articulation}
  \label{fig:11}
\end{subfigure}
\caption{Average Topographic map of the prompt "$/n/$"}
  \label{fig:topoerp}
  \vspace{-0.3cm}
\end{figure}

\vspace{-0.1cm}
\section{Conclusion}\label{sec:conc}
Employing correlation networks to model inter-phasal information and using the same  for better decoding of imagined phase EEG is discussed in this work. Two levels of feature extraction are designed, a primary one using handcrafted features and a secondary one using CorNet-derived features. Popular machine learning and deep learning models are then used to classify the EEG units. The accuracies reported using the proposed approach are higher than previously reported phonological classification performances using the KARA-ONE database. The work further recommends an efficient way of handling the multi-phasal/modal data by proposing a single-phase decode approach. Results show that from a correlation perspective, the speaking phase EEG contains more supportive information as compared to articulatory EEG. 
Further, visualizations using topographic maps provide evidence of correlation between the phases that can be exploited as is done in the proposed approach for improved imagined EEG unit classification.

\clearpage
\bibliographystyle{IEEEtran}
\bibliography{mybib}

\end{document}